\documentclass[prl,twocolumn,floatfix,superscriptaddress]{revtex4}

\usepackage{epsf}
\usepackage{graphicx}
\usepackage{dcolumn}
\usepackage{bm}

\newcommand{\be}{\begin{equation}}
\newcommand{\ee}{\end{equation}}
\newcommand{\ba}{\begin{eqnarray}}
\newcommand{\ea}{\end{eqnarray}}
\newcommand{\nn}{\nonumber \\}

\begin{document}

\title{Violation of the fluctuation-dissipation theorem in time-dependent mesoscopic heat transport}

\author{Dmitri V.~Averin}
\affiliation{Department of Physics and Astronomy, Stony Brook
University, SUNY, Stony Brook, NY 11794-3800 }

\author{Jukka P. Pekola}

\affiliation{Low Temperature Laboratory,  Aalto University,
P.O.~Box 13500, FI-00076 AALTO, Finland}

\date{\today}

\begin{abstract}
We have analyzed the spectral density of fluctuations of the energy flux through a mesoscopic constriction between two equilibrium reservoirs. It is shown that at finite frequencies, the fluctuating energy flux is not related to the thermal conductance of the constriction by the standard fluctuation-dissipation theorem, but contains additional noise. The main physical consequence of this extra noise is that the fluctuations do not vanish at zero temperature together with the vanishing thermal conductance.

\end{abstract}

\maketitle

Fluctuation-dissipation theorem (FDT) \cite{cw} relates the fluctuations of a dynamic variable generated at angular frequency $\omega$ by an equilibrium  statistical-mechanical system, to the dissipative part of the response function of this system to the force conjugate to this variable at the same frequency. This theorem is one of the main physical predictions of the ``linear response theory'' of equilibrium transport properties in  statistical mechanics, and finds applications in practically all areas of the condensed-matter physics. The best known case of this theorem is the relation between the electric conductance of a resistor expressed through Kubo formula \cite{kubo} and the current noise generated by this resistor. This relation played important role in understanding macroscopic quantum dynamics of superconducting structures, where the Josephson effect provides a way of directly observing the quantum part of current noise as predicted by the FDT \cite{koch}. More recently, this theorem found applications within the studies of the mechanisms of decoherence and noise in superconducting qubits (see, e.g., \cite{mrt}). From the perspective of the linear-response theory, thermal transport represents a somewhat special case, since temperature $T$ (more precisely, the temperature gradient) that acts as the force conjugate to the heat current, does not correspond microscopically to any dynamic degree of freedom of a statistical system.  Nevertheless, it is frequently assumed that the FDT holds also for heat transport, and relates the spectral density of the energy current $S(\omega)$ and the heat conductance $G_{th}(\omega)$ by (see, e.g., \cite{ll})
\be S (\omega) = \hbar \omega \, T \mbox{Re}\,  G_{th}(\omega) \coth (\hbar \omega/2T) \, , \label{e1} \ee
This assumption is supported by the fact that at zero frequency, Eq.~(\ref{e1}) reduces to the form
\be S (0) = 2\, T^2 G_{th}(0)\, , \label{e2} \ee
that follows directly from the fundamental thermodynamic result for the magnitude of energy fluctuations. Here and everywhere below we define temperature $T$ in energy units, by setting $k_B=1$.

\begin{figure}[ht]\center
\includegraphics[width=0.6\linewidth]{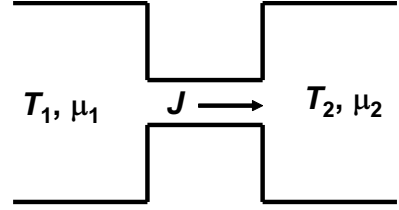}
\caption{\label{fig1} Schematics of a generic constriction admitting energy current $J$ between two equilibrium reservoirs with temperatures $T_j$ and chemical potentials $\mu_j$, $j=1,2$.}
\end{figure}

The purpose of this work is to demonstrate explicitly that the thermal FDT of Eq. (\ref{e1}) is not valid at non-vanishing frequencies, at least as a general statement. To do this, we calculate the spectral density $S (\omega)$ of the fluctuations $\tilde{J}=J-\langle J\rangle$ of the energy current $J$:
\be S (\omega) = \int d t e^{-i\omega t} [\tilde{J}(t)\tilde{J}(0) +\tilde{J}(0) \tilde{J}(t)]/2 \, , \label{e3} \ee
in the set-up characteristic for a ``mesoscopic'' measurement of heat transport, see Fig. \ref{fig1}. We consider two reservoirs, equilibrated to, in general, different temperatures $T_{1,2}$, and weakly coupled by a constriction that supports propagation of excitations that carry energy between the reservoirs. Such a general set-up describes both the heat transport by phonons, e.g. in demonstrations of the quantization of the phonon heat conductance \cite{ph1}, and also various structures of ``on-chip cryogenics'' \cite{rmp} where the heat is transported by electrons or photons \cite{schmidt,meschke}. The special status of the temperature $T$ as a parameter in the density matrix of the system, instead of being a dynamic variable, creates a problem for the microscopic definition of thermal conductance. The system Hamiltonian should have the property of maintaining local thermal equilibrium required for the temperature and the corresponding thermal bias to be well-defined, while at the same time creating the non-vanishing energy flux through the system (see, e.g., the discussion in \cite{pba}). An important feature of the mesoscopic set-up considered here (Fig.~1), is that this problem is resolved naturally by clear separation of the two processes. Temperature-defining equilibrium is maintained inside the reservoirs, while the energy flux is created by weak perturbative coupling between them.

Quantitatively, we first treat the case of phonon heat transport. (With some minor modifications, the same considerations and results apply, obviously, to the photon heat conduction.) We are interested in the regime of frequencies/energies small on the scale set by the phonon propagation time along the constriction, $\omega, T/\hbar \ll 1/\tau_{tr}$, where $\tau_{tr}=l/v$, with the sound velocity $v$ and constriction length $l$. In this low-frequency limit, the short constriction does not lead to any reflection resulting in ballistic propagation of phonons. The simplest description of the energy transport by such ballistic phonons in one traversal mode starts with the usual Hamiltonian $H$ of the field  $\phi(x)$ of the longitudinal one-dimensional (1D) phonons:
\be
H=\int dx h(x)\, , \;\;\; h(x)=\frac{1}{2} \{ \rho [\dot{\phi}(x)]^2 +\kappa [\phi'(x)]^2 \}\, ,  \label {e6} \ee
where $\rho$ and $\kappa$ are the density and compressibility of the constriction, so that $v=(\kappa/\rho)^{1/2}$. Writing the Heisenberg equation of motion for the energy density $h$ in the form of the continuity equation, $\dot{h}+J'=0$, one obtains the following expression for the operator of the energy flux $J$ carried by phonons (see, e.g., \cite{lutt}):
\[ J=-\frac{\kappa}{2} [\dot{\phi} \, \phi'+ \phi' \dot{\phi} ] \, .  \]
The usual mode expansion of the field $\phi$,
\[ \phi(x)=\big(\frac{\hbar}{2L \rho}\big)^{1/2} \sum_k (a_ke^{ikx}+h.c.)/\omega_k ^{1/2}, \]
where $\omega_k=v|k|$ and $L$ is a normalization length, gives
\be
J= \frac{\hbar v}{2L} \sum_{k,p} (\omega_k \omega_p)^{1/2} \mbox{sign}(k) (a_k-a_k^{\dagger})(a_p^{\dagger}-a_p) \, . \label {e4} \ee
(Since none of the quantities we consider below depend on $x$ in the small-$\tau_{tr}$ limit after thermal averaging, we set $x=0$ in this expression.) Averaging $J$ over the equilibrium states of the phonon modes, and taking the appropriate limit $L\rightarrow \infty$, one obtains the average energy current:
\be
\langle J\rangle = \int_0^{\infty} \frac{dE E}{2\pi \hbar } [n_1(E)- n_2(E)]= \frac{\pi }{12\hbar } [T_1^2-T_2^2] \, , \label {e8} \ee
where $n_j(E)$ is the Bose distribution at temperature $T_j$. Expansion of
Eq.~(\ref{e8}) in small temperature difference $\delta T/2$: $T_{1,2}=T\pm \delta T/2$, gives the standard expression for the ``quantum'' of the heat conductance of one phonon channel, $G_{th}=\pi T/6 \hbar$. The point of our derivation here is that Eq.~(\ref{e8}) remains valid even if the temperatures $T_{1,2}$ of the phonons incident on the constriction vary in time with frequencies less than $1/\tau_{tr}$. This means that the resulting expression for the heat conductance $G_{th}$ is also valid for all frequencies in this range.

Next, substituting the operator $J$ given by Eq.~(\ref{e4}) into Eq.~(\ref{e3}) and repeating the same steps that lead to the average energy current (\ref{e8}), we obtain after some algebra the following expression for the spectral density of the energy flux noise:
\[ S (\omega) = \frac{1}{8\pi \hbar }\sum_{\pm, \, j} \int dE E(E\pm \Omega) (1+n_j(E))n_j(E\pm \Omega) \, , \]
$\Omega \equiv \hbar \omega$. This expression generalizes to finite frequencies previous results for the noise in phonon heat transport (see, e.g., \cite{ph2}). Taking the integral, we get
\be S (\omega) = \frac{1}{48 \pi \hbar} \sum_j [(2\pi T_j)^2+\Omega^2] \Omega \coth \frac{\Omega}{2T_j} \, . \label{e10} \ee
One can see that even in equilibrium, $T_1=T_2\equiv T$, the spectral density (\ref{e10}) is different at $\Omega \neq 0$ from the one predicted by the FDT, see Eq.~(\ref{e1}), with the conductance $G_{th}(\omega)=\pi T/6 \hbar$. The FDT reproduces only the first part of Eq.~(\ref{e10}) that corresponds to the $T^2$-term in the brackets. In addition to this, the full result (\ref{e10}) contains the $\Omega^2$-term that is non-vanishing, $S (\omega) = \hbar^2 |\omega|^3/(24 \pi)$, even at $T=0$, when the heat conductance is zero. Physically, the origin of this extra term can be traced back to finite coupling between the reservoirs, which creates quantum fluctuations of their energy even at $T=0$, when the thermal conductance in the FDT relation (\ref{e1}) vanishes since there are no real excitations that could irreversibly transfer energy between the reservoirs. In this respect, the violation of the FDT for the heat transport discussed in this work has the same origin as several other thermodynamic effects of finite relaxation energy that have been discussed in the literature \cite{b7,b8,b9,b10}.

This qualitative picture implies that the breakdown of the FDT for heat transport is not a specific feature of the phonon heat conduction, but is quite general. To demonstrate this, we consider a similar set-up of two weakly-coupled reservoirs but in the situation when the heat conductance  is due to electron propagation between them. The main difference with the phonon case is that the reflection in the constriction can be non-negligible for electrons even when the traversal time $\tau_{tr}$ is very short on the scale set by other energies in the problem. The calculation of the finite-frequency heat transport by electrons follows the same steps as in the case of phonons. The operator of energy density $h(x)$ of 1D electrons can be written in terms of the electron field $\psi(x)$ and the single-particle Hamiltonian $\hat{h} =-(\hbar^2/2m)\partial^2/\partial x^2 +V(x)$ as
\be h(x) =  [\psi^{\dagger} \hat{h} \psi  +(\hat{h} \psi^{\dagger}) \psi]/2 \, . \label{e11} \ee
The symmetrized expression (\ref{e11}) is needed to ensure that $h(x)$  is Hermitian. Then, the Heisenberg equation of motion for $h(x)$ with the Hamiltonian $H=\int dx h(x)$ takes the form of the continuity equation, $\dot{h}+J'=0$, with the energy flux operator
\be J= (- i\hbar/4m) [\psi^{\dagger} (\hat{h} \psi)' - (\psi^{\dagger})' \hat{h} \psi -h.c.] \, . \label{e12} \ee
This expression shows that if one decomposes the field $\psi$ into the stationary scattering modes, the energy current has the same form as the usual probability current, the only difference being that $\hat{h}$ multiplies each mode by its energy. Explicitly, introducing the creation/annihilation amplitudes $a_k^{\dagger}, a_k$ for electrons incident from one electrode, and  $b_k^{\dagger}, b_k$ from the other, we obtain the following mode expansion of $J$:
\be J= \frac{v_F}{L} \sum_{k,p} \frac{\epsilon_k+ \epsilon_p}{2} [D(a_k^{\dagger} a_p-b_k^{\dagger}b_p)+\sqrt{DR}(a_k^{\dagger} b_p+b_k^{\dagger}a_p)]  \, . \label{e13} \ee
Here $v_F$ is the Fermi velocity, $D$ and $R$ are, respectively, the transmission and reflection probabilities of the constriction, $D+R=1$, and $\epsilon_{k,p}$ are the electron energies. Also, we assume that both the Fermi energies in the electrodes, and $\hbar/\tau_{tr}$ are much larger than the typical excitation energies $T$, $\Omega$, and $eV$, where $V$ is the bias voltage between the electrodes. This implies, in particular, that the scattering probabilities $D,R$ are constant in the energy range of interest.
The energy current $J$ (\ref{e13}) corresponds directly to the heat flow into/out of the reservoir $j$, if the electron energies $\epsilon_{k,p}$ are measured in Eq.~(\ref{e13}) relative to the chemical potential $\mu_j$ of this reservoir. (Note that in the phonon calculation above, this condition was satisfied automatically, since $\mu=0$ for phonons.) If $V=0$, so that there is no shift between the chemical potentials of the two electrodes, the average of Eq.~(\ref{e13}) represents both the heat flow, $-J_1$, out of one electrode and the heat flow, $J_2$, into the other one: $-\langle J_1 \rangle = \langle J_2\rangle =\langle J\rangle$. The total average generated heat is zero, $ \langle J_1+J_2 \rangle =0$. If, however, $V\neq 0$, then one needs to measure the energies relative to the two different levels $\mu_1$ and $\mu_2$ in the two reservoirs, and $-\langle J_1 \rangle \neq \langle J_2\rangle $. The difference between the two heat flows is obtained by replacing the energy $(\epsilon_k+ \epsilon_p)/2$ in Eq.~(\ref{e13}) with $\mu_1-\mu_2=eV$. After this substitution, Eq.~(\ref{e13}) reduces to $IV$, where $I$ is the electric current between the reservoirs. Therefore, in the case of non-vanishing bias voltage $V$, the total generated heat is non-vanishing, and equal to the Joule heat, $\langle J_1+J_2\rangle =\langle I\rangle \, V$, where the individual heat flows $J_j$ are obtained from Eq.~(\ref{e13}) by measuring the energies $\epsilon_{k,p}$ relative to the chemical potential $\mu_j$ of the corresponding electrode.

Taking thermal average and the limit $L\rightarrow \infty$ in Eq.~(\ref{e13}), we find the average heat currents $\langle J_j \rangle$, $j=1,2$, into the two electrodes:
\[ \langle J_j\rangle =(-1)^j \frac{D}{2\pi \hbar } \int dE (E-\mu_j) [f_1(E)- f_2(E)] \, , \]
where $f_j(E)$ is the Fermi distribution function of electrons in the $j$th reservoir. This gives for the heat currents:
\be
\langle J_j\rangle = \frac{D}{2\pi \hbar }[\frac{(eV)^2}{2} +(-1)^j\frac{\pi^2}{6} (T_1^2-T_2^2)] \, . \label {e14} \ee
The two terms in this expression represent, respectively, the usual Joule heating, which in this case is distributed equally between the electrodes, and the heat transport between them. For small temperature difference between the electrodes, Eq.~(\ref{e14}) gives the thermal conductance, $G_{th}=\pi DT/6\hbar$, that coincides with the phonon conductance. As with the phonons, an important point of our derivation here is that this thermal conductance is independent of frequency in the considered frequency range below $1/\tau_{tr}$ and the frequencies set by the Fermi energies in the electrodes, and the energy scale of the variations of the transmission  probability.

At $V\neq 0$, when the total heat fluxes into the two electrodes are different due to Joule heating, to describe specifically the heat transfer between the electrodes, one needs to define the heat current as $J=(J_2-J_1)/2$. Indeed, as one can see from Eq.~(\ref{e14}), the average heat current $\langle J\rangle$ defined this way is not affected by the Joule heating. This definition corresponds to the simple prescription of measuring all energies in Eq.~(\ref{e13}) relative to the midpoint between the chemical potentials of the two electrodes. Then, the same steps as for the average current, including thermal averaging and the $L\rightarrow \infty$
limit, give for the spectral density of the heat current noise
\ba  S (\omega) = \frac{D}{4\pi \hbar }\sum_{\pm,\, j} \int dE
(E\pm \Omega/2)^2\big[D f_j(E)\cdot \nn (1-f_j(E\pm \Omega)) +R f_j(E)(1+f_{j'}(E\pm \Omega)) \big] \, , \label{e15} \ea
where $j'$ is defined as $j' \neq j$ with $j, j'=1,2$. One can see that without the bias voltage and reflection, $V=0$, $R=0$, Eq.~(\ref{e15}) coincides with the phonon result (\ref{e10}). For equal temperatures of the electrodes, $T_1=T_2$, calculation of the integral (\ref{e15}) gives
\ba S (\omega) = \frac{D}{48 \pi \hbar} \big\{2D[\Omega^2+(2\pi T)^2+3(eV)^2] \Omega \coth \frac{\Omega}{2T} \nn
+ R \sum_{\pm}[(eV\pm \Omega )^2+(2\pi T)^2] (eV\pm \Omega )\coth
\frac{eV \pm \Omega}{2T}\big\} . \;\; \label{e16} \ea

This result (as well as Eq.~(\ref{e10}) for the phonons) can be extended naturally to the situation when the contact between the electrodes supports many electron modes with transparencies $D_k$. In the equilibrium case that is of the main interest here, Eq.~(\ref{e16}) gives the following expression for the heat noise in such a multi-mode contact:
\be S (\omega) = (G/12 e^2) [\Omega^2+ (2\pi T)^2] \Omega \coth (\Omega/2T) \, . \label{e17} \ee
Here $G=\sum_k D_k e^2/(2\pi \hbar)$ is the electric conductance of the
contact, which is related by the Wiedemann-Franz law to the heat conductance  $G_{th}= \pi^2 GT/(3e^2)$.

One of the interesting physics features of Eq.~(\ref{e16}) in the case of  non-vanishing $V$ and $R$ is the shot noise of the energy current associated with the individual electron scattering events. Energy currents carried by individual electrons were predicted in \cite{jp07} and demonstrated recently in the form of RF-cooling in a metallic single-electron transistor with an alternating voltage at the gate \cite{sg09}. Equation (\ref{e16}) shows that in a biased contact, the average heat current is accompanied by shot noise of heat due to the scattering of discrete electrons. Quantitatively, the same multi-mode generalization of (\ref{e16}) gives this noise for $|eV| \gg T,\Omega$ as
\be S (V) = e F G|V|^3/12\, , \label{e18} \ee
where $F=\sum_k D_k(1-D_k)/\sum_k D_k$ is the standard Fano factor that characterizes the shot noise of electric current.

Returning to Eq.~(\ref{e17}) and quantum fluctuations of heat, we see again that, similarly to the situation with the phonon heat conductance, equilibrium fluctuations of the heat current are not described correctly by FDT at finite frequencies. While the heat conductance of the contact is frequency-independent in the range discussed above, the noise contains frequency-dependent part which does not vanish at $T=0$ together with the heat conductance. Physically, these fluctuations are produced by virtual electron transitions between the two electrodes due to finite coupling between them. This mechanism is the same as for the quantum fluctuations of electrical current in the contact, and Eq.~(\ref{e16}) for the energy fluctuations is quite similar to the corresponding equation for the current fluctuations. The fact that the quantum fluctuations of electric current are still consistent with the FDT, in particular, vanish together with the electric conductance $G$, while the thermal fluctuations do not agree with FDT, is a reflection of the special nature of temperature in statistical mechanics.

As the last point of our discussion of the thermal FDT, we would like to make more explicit the set of assumptions underlying the notion of the frequency-dependent thermal conductance $G_{th} (\omega)$. We do this in the case of electron heat transport considered above, limiting the discussion to the tunnel approximation $D\ll 1$, when electron scattering can be described with the usual tunnel Hamiltonian
\be
H=H_1+H_2+H_T\, , \;\;\; H_T= \sum_{k,p} (T_{kp}a_k^{\dagger} b_p+ h.c.) \, ,\label{e19} \ee
which explicitly separates Hamiltonian of the electrodes $H_{1,2}$ and the tunneling term $H_T$. Here $T_{kp}$ are the tunneling amplitudes that can be expressed through the junction conductance $G$. The heat flow defining the thermal conductance at finite frequencies should be driven by a small time-dependent temperature difference $\delta T(t)$. Separating one frequency $\omega$, we take $T_{1,2}=T\pm (\delta T/2)e^{-i\omega t}$. Expansion of the density matrix of the equilibrium electrodes in $\delta T \ll T$ gives the $\delta T$-induced correction $\delta \rho$ to it as
\be
\delta \rho (t) =- \rho_0 Q (\delta T/2T^2) e^{-i\omega t}. \label{e21} \ee
Here $\rho_0=(1/Z) e^{-(H_1+H_2)/T}$ is the main part of the density matrix and $Q= Q' -\langle Q'\rangle$, with $Q'\equiv H_2-H_1$ and $\langle ... \rangle$ denoting the average over $\rho_0$.  We note that the assumption of equilibrium with the time-dependent temperature requires that $\omega \ll 1/\tau_E$, where $\tau_E$ is the relaxation time in the electrodes. This condition is not satisfied directly in the model of non-interacting phonons or electrons for which we calculated the energy flux noise $S(\omega)$. One can still use Eq.~(\ref{e21}) for the non-interacting particles incident on the junction, assuming that the temperature-defining relaxation is concentrated infinitely deep inside the electrodes. The usual perturbation theory in $H_T$ around $\delta \rho$ (\ref{e21}) then gives
\be
G_{th} (\omega) = \frac{i}{2\hbar T^2} \int_0^{\infty} dt e^{i\omega t}
\langle J(t) H_T Q -QH_T J(t) \rangle ,\label{e22} \ee
where the energy flux operator is
\[ J=\dot{Q}/2= (i/2\hbar) \sum_{k,p} (\epsilon_k+ \epsilon_p) (T_{kp} a_k^{\dagger} b_p- h.c.) \, . \]

Equation for the thermal conductance similar to (\ref{e22}) can also be derived in a general situation, when the heat current is driven by some arbitrary relaxation interaction $V$ (and not the tunneling $H_T$). Formally, the break-down of the FDT for the heat transport discussed in this work arises from the difference between the structure of Eq.~(\ref{e22}) and expressions for the ``dynamic'' linear-response coefficients, e.g., the electric conductance. Explicitly, evaluating (\ref{e22}) we get $\mbox{Re}\, G_{th} (\omega) = \pi^2 GT/(3e^2)$, which is frequency-independent in agreement with the arguments provided above, and in contradiction to the FDT (\ref{e1}), if compared with the energy flux noise (\ref{e17}).

We would like to thank P.B. Allen, T.T. Heikkil\"{a}, N.B. Kopnin, and K.K. Likharev for useful discussions. This work was supported in part by the Academy of Finland.

\end{document}